\documentclass[11pt,twoside]{article}
\usepackage{asp2010}
\usepackage{graphicx}

\resetcounters

\begin{document}

\title{Emerging Flux Simulations and Proto-Active Regions}
\author{Robert~F.~Stein$^1$, Anders~Lagerfj{\"a}rd$^2$, {\AA}ke~Nordlund$^2$ and Dali Georgobiani$^1$}
\affil{$^1$Michigan State University, East Lansing, MI 48824, USA}
\affil{$^2$Niels Bohr Institute, Blegdamsvej 17, 2100 K{\/o}benhavn {\O}, DK}

\begin{abstract}
The emergence of minimally structured (uniform and horizontal)
magnetic field from a depth of 20 Mm has been simulated.  The field
emerges first in a mixed polarity pepper and salt pattern, but then
collects into separate, unipolar concentrations and produces pores.
The field strength was then artificially increased to produce
spot-like structures.   The field strength at continuum optical
depth unity peaks at 1 kG, with a maximum of 4 kG.  Where the
vertical field is strong, the spots persist (at present an hour of
solar time has been simulated).  Where the field is weak, the spot
gets filled in and disappears.  Stokes profiles have been calculated
and processed with the Hinode annular mtf, the slit diffraction and
frequency smoothing.  These data are available at
steinr.pa.msu.edu/$\sim$bob/stokes.
\end{abstract}

\section{Introduction}

The rise of a coherent, straight, horizontal, twisted flux tube through the shallow layers
of realistic solar convection zones has been studied by
\cite{Cheung08,Cheung07,Cheung06} and \citet{Martinez-Sykora09,Martinez-Sykora08}.
The structure of sunspots has been simulated by \citet{Rempel09a,Rempel09b}
starting from an initial state of a coherent, vertical, cylindrical magnetic
flux tube and by \citet{Cheung10} for a buoyant, coherent, twisted,
semi-torus flux tube kinematically advected into the domain from a
depth of 7.5 Mm.  We have investigated a complementary configuration
where minimally structured, uniform, untwisted, horizontal field
is advected into the computational domain by convective inflows
through the bottom at a depth of 20 Mm.  Earlier results were
reported by \citet{Stein10}.

\section{The Simulation}

We use the ``Stagger-Code", which solves the equations for mass,
momentum and internal energy in conservative form plus the induction
equation for the magnetic field, for fully compressible flow, in
three dimensions, on a staggered mesh.  The dimensions are 48 Mm
wide and extends from the temperature minimum at the top of the
photosphere to a depth of 20 Mm (half the scale heights of the
convection zone).  The code uses finite differences, with 6th order
derivative operators and 5th order interpolation operators.  The
grid is uniform and periodic in horizontal directions and non-uniform
in the vertical (stratified) direction.  Time integration is by a
3rd order low memory Runge-Kutta scheme \citep{Kennedy99}.
Parallelization is achieved with MPI, communicating the three overlap
zones that are needed in the 6th and 5th order derivative and
interpolation stencils.   Horizontal directions are periodic.  Top
and bottom boundaries are transmitting.  Inflows at the bottom
boundary have uniform pressure, specified entropy and damped
horizontal velocities.  Outflow boundary values are obtained by
extrapolation.  The magnetic field is made to tend toward a potential
field at the top and at the bottom is given a specified value in
inflows and extrapolated in outflows.

We use a tabular equation of state that includes local thermodynamic
equilibrium (LTE) ionization of the abundant elements as well as
hydrogen molecule formation, to obtain the pressure and temperature
as a function of log density and internal energy per unit mass.  We
calculate the radiative heat/cooling by solving the radiation
transfer equation in both continua and lines using the Feautrier
method \citep{Feautrier64}, assuming Local Thermodynamic Equilibrium
(LTE).  The number of wavelengths for which the transfer equation
is solved is drastically reduced by using a multi-group method
whereby the opacity at each wavelength is placed into one of four
bins according to its magnitude and the source function is binned
the same way \citep{Nordlund82,Stein+Nordlund_3drad02,Vogler04}.

At supergranule and larger scales, the coriolis force of the solar
rotation begins to influence the plasma motions, so f-plane rotation
is included in the simulation.  Angular momentum conservation
produces a surface shear layer with the surface (top of the domain)
rotating slower than the bottom of the domain, as observed in the
Sun.

The simulation is started from a snapshot of hydrodynamic convection
and inflows at the bottom (20 Mm depth) advect in uniform, untwisted,
horizontal field.

\section{Emerging Flux}

\begin{figure}
  \centerline{\includegraphics[width=\textwidth]{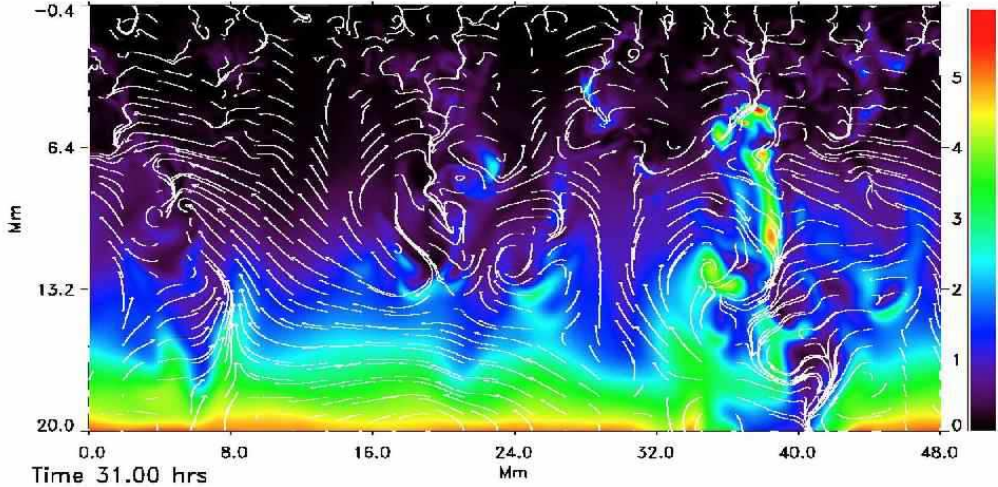}}
  \caption{Magnetic field (color) advected through the bottom
  boundary by inflows, rising through the upper convection zone.  The 
  plasma flow (white arrows) advects the field toward the surface in 
  upflows and inhibits its rise in downflows.  Scale is in kG.
  }
  \label{fig:risingB}
\end{figure}
  
\begin{figure}[!htb]
  \centerline{\includegraphics[width=.9\textwidth]{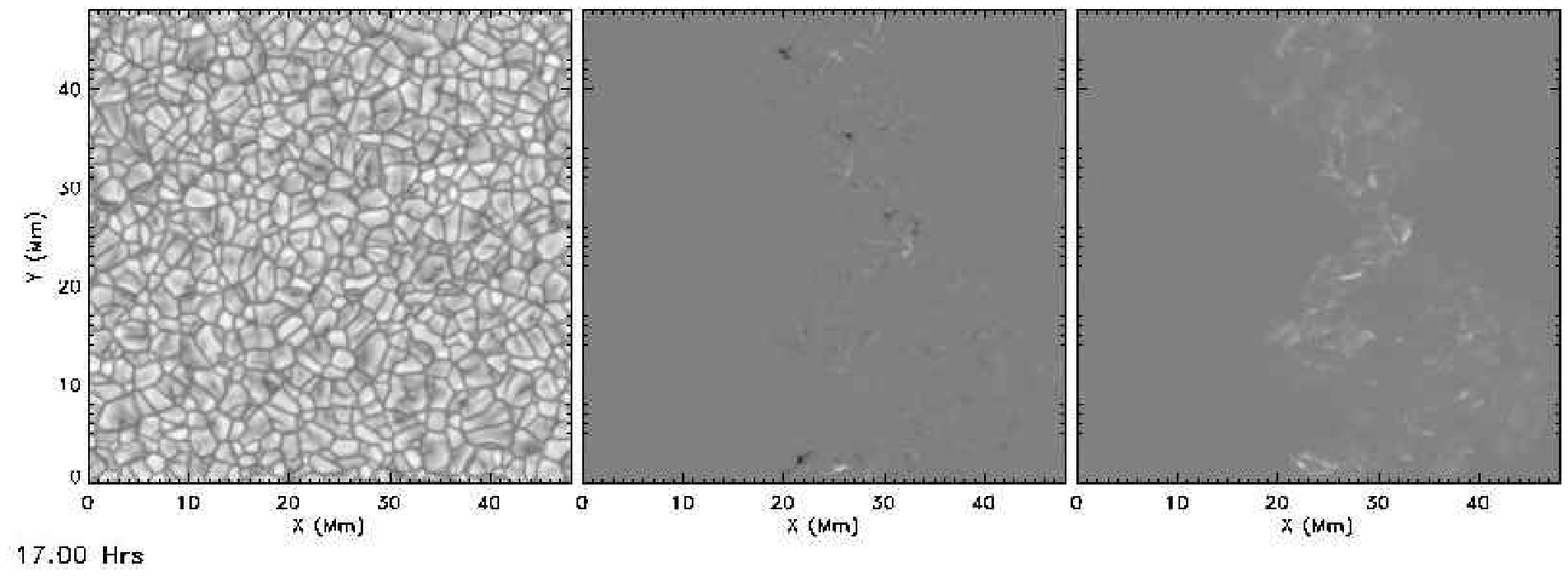}}
  \centerline{\includegraphics[width=.9\textwidth]{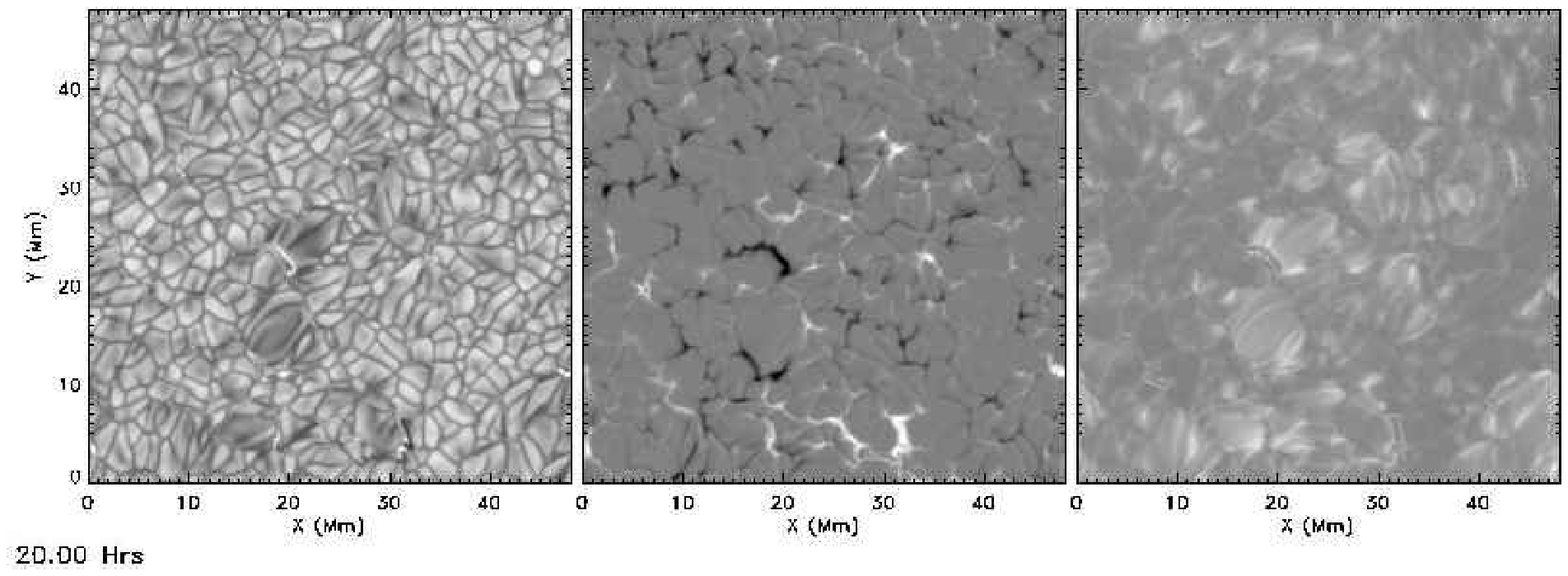}}
  \centerline{\includegraphics[width=.9\textwidth]{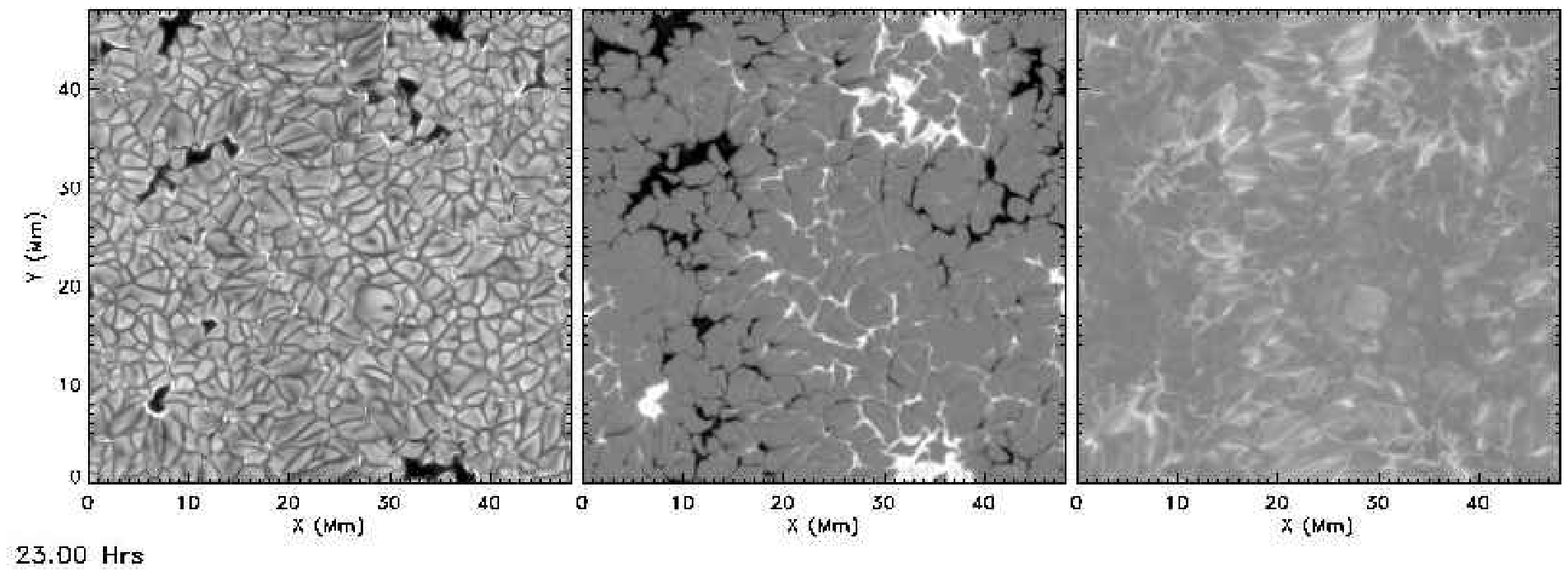}}
  \centerline{\includegraphics[width=.9\textwidth]{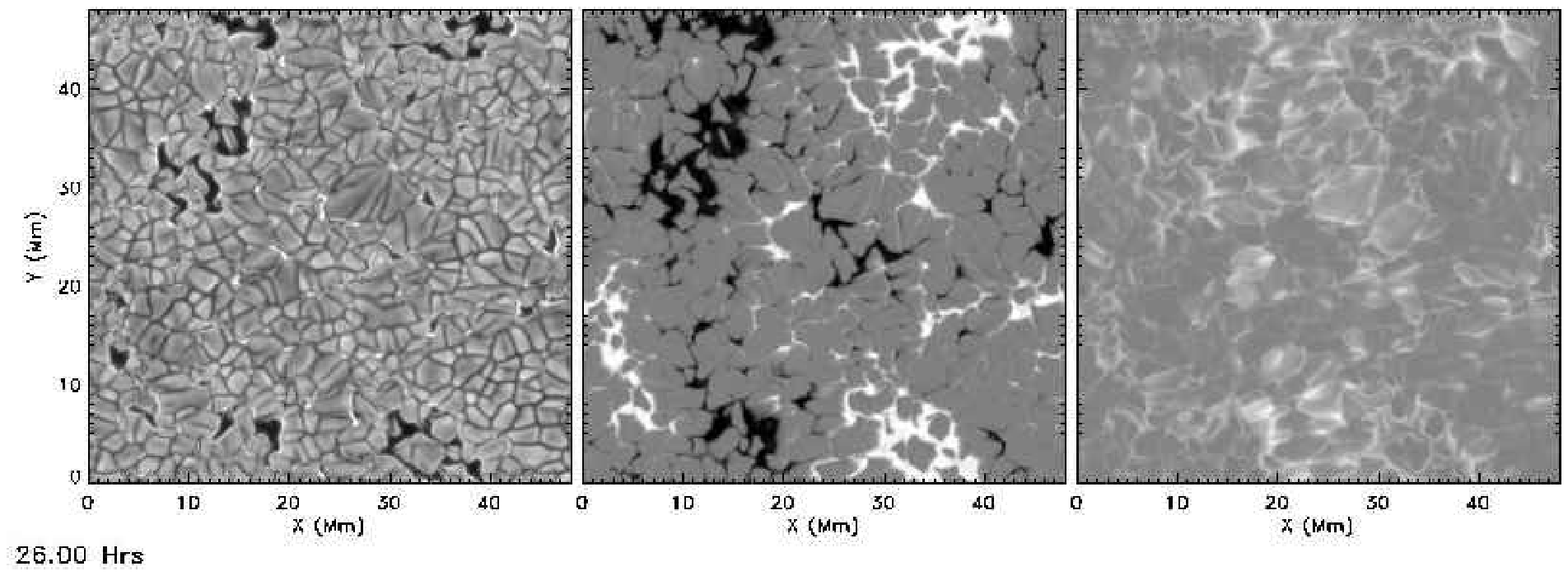}}
  \caption{Emergent continuum intensity (left), vertical magnetic
  field at $\tau_{\rm cont}=0.01$ (center) and horizontal magnetic
  field at the same optical depth (right).  The range of intensities
  is 0.22-1.35$<I>$.  The range of magnetic field is $\pm 2$ kG.  Time is
  since horizontal field started entering at 20 Mm depth.  The typical 
  fluid rise time is 32 hours from that depth.
  }
  \label{fig:bsurfseq}
\end{figure}
 
\begin{figure}
  \centerline{\includegraphics[width=13cm,height=6.5cm]{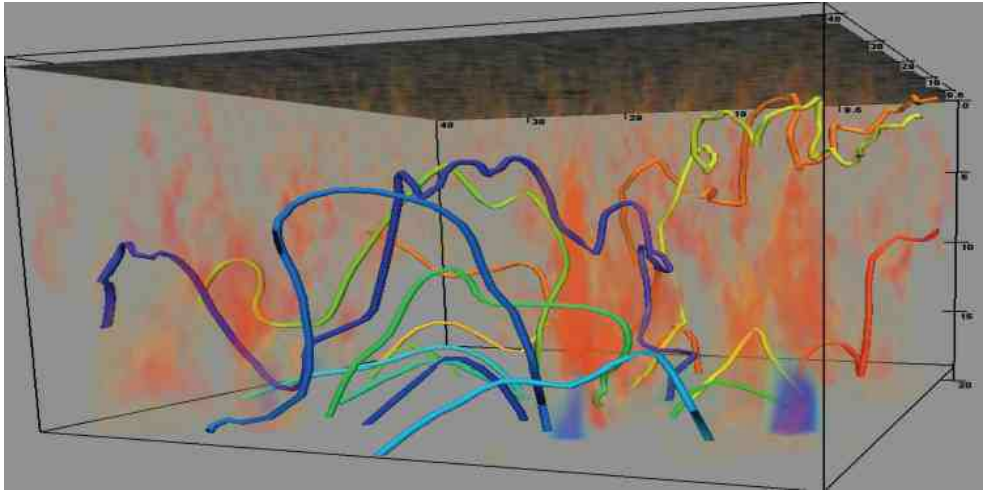}}
  \caption{Several magnetic field lines show large scale loops with
  smaller serpentine loops riding piggy-back on them.   Shading
  shows downflows.
  }
  \label{fig:blines}
\end{figure}

Upflows, aided by magnetic buoyancy, carry the magnetic flux to the
surface.  Downflows tend to pin the field down (Fig.~\ref{fig:risingB}).
This results in the development of magnetic loops.  The typical
time for fluid to rise from a depth of 20 Mm to the surface is 32
hours.  In the present case of 20 kG field at 20 Mm depth, the
magnetic flux reaches the surface in approximately 20 hours.  Magnetic
flux initially appears in a small localized portion of the surface,
but quickly spreads to cover the entire surface in a pepper and
salt pattern of mixed polarity.  The different polarities then
collect into unipolar regions (Fig.~\ref{fig:bsurfseq}).  The small
scale convective motions near the surface produce small serpentine
loops that ride piggy-back on the larger loops produced by supergranule
scale convective motions at large depths (Fig.~\ref{fig:blines}).

\section{Active Region}

\begin{figure}
  \centerline{\includegraphics[width=.9\textwidth]{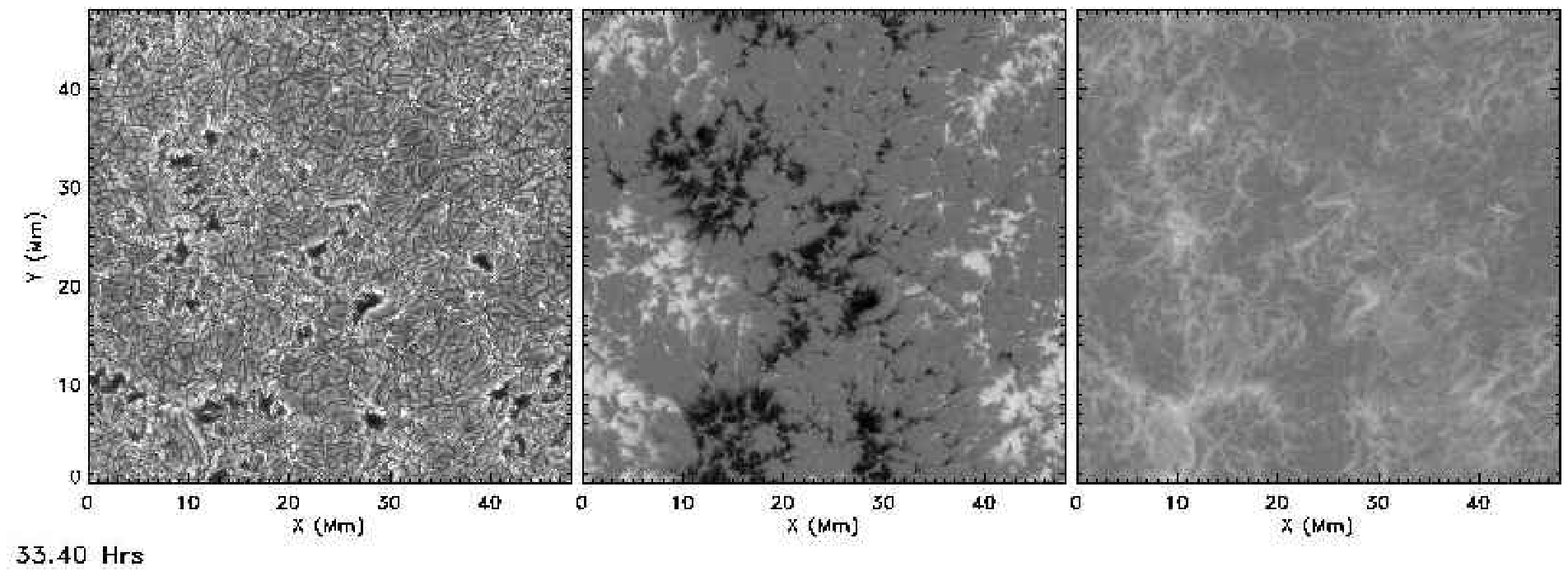}}
  \centerline{\includegraphics[width=.9\textwidth]{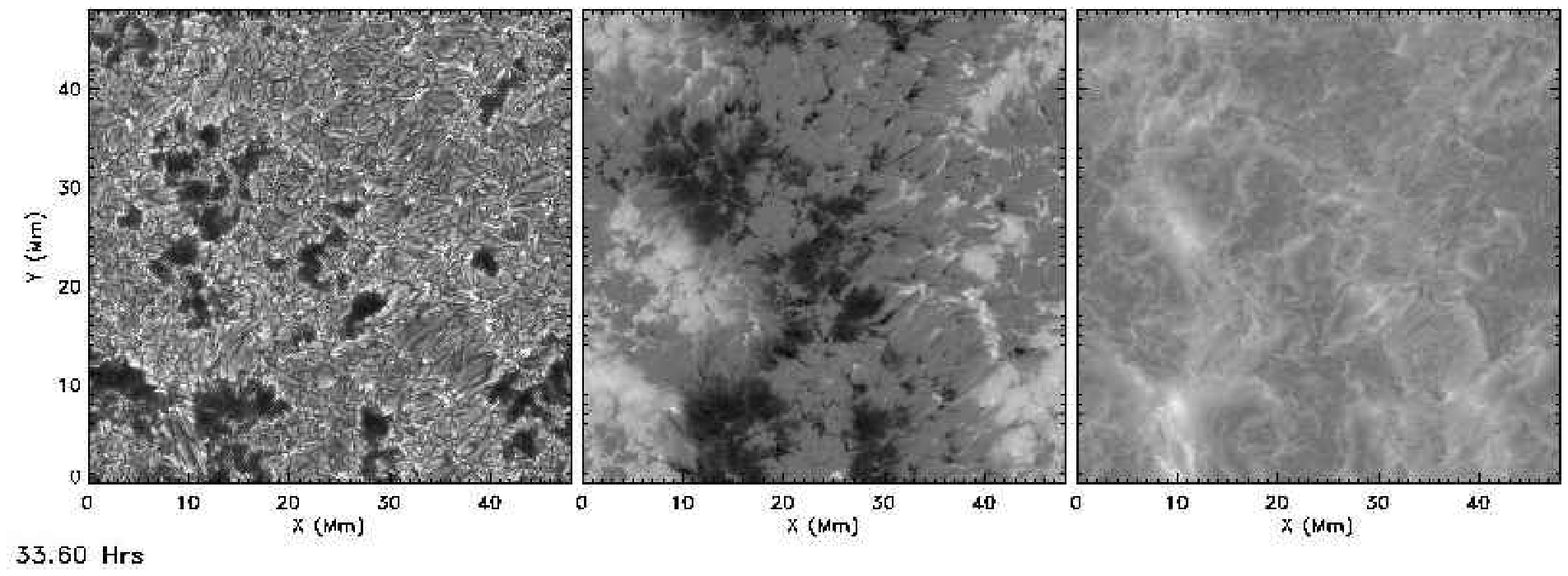}}
  \centerline{\includegraphics[width=.9\textwidth]{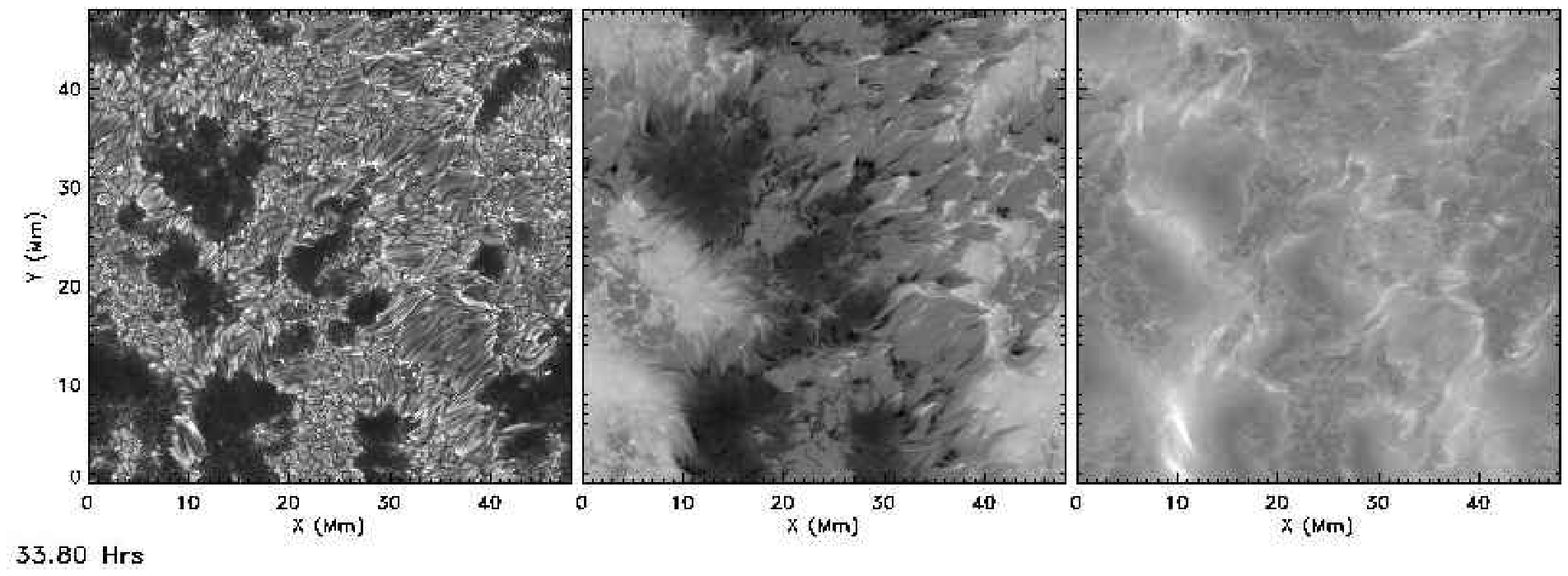}}
  \centerline{\includegraphics[width=.9\textwidth]{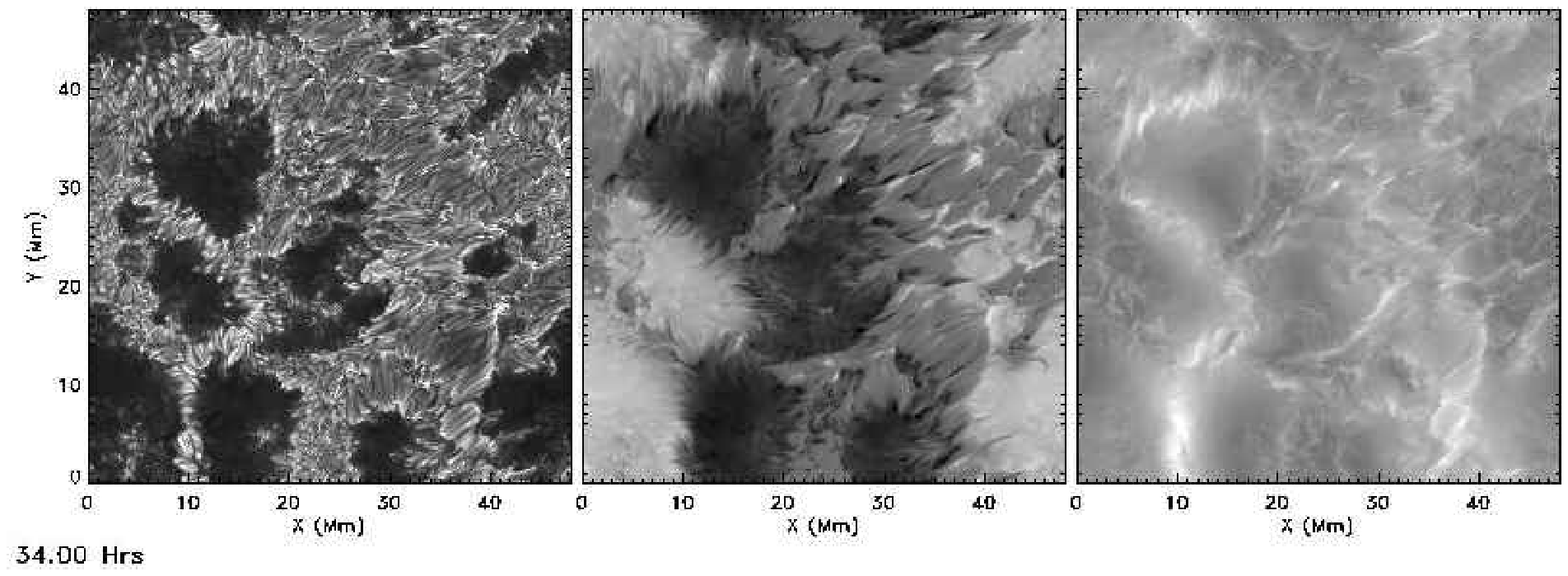}}
  \caption{Time sequence showing how the active region is produced
  by increasing the magnetic field strength $\propto B$.  Emergent
  continuum intensity (left), vertical magnetic field at $\tau_{\rm
  cont}=0.01$ (center) and horizontal magnetic field at the same
  optical depth (right).  The range of intensities is 0.15-2$<I>$.
  The range of magnetic field is $\pm 3$ kG.  Time is since the
  start of horizontal field being advected into the 20 Mm deep
  domain.
  }
  \label{fig:ARgrowth}
\end{figure}

Once sufficient magnetic flux had arrived at the surface
(Fig.~\ref{fig:bsurfseq}), we artificially increased the magnetic
field strength, in proportion to its existing strength, with a time
scale of 30 minutes (Fig.~\ref{fig:ARgrowth}).  This made the area
occupied by the strong fields and the pores they had produced increase.
However, the field strength at a fixed optical depth increased only 
slightly as the Wilson depression deepened and the surrounding
gas pressure at fixed optical depth in the flux concentration increased.  
This method of simulating an active region
allows the magnetic structures to develop naturally in magneto-convection
before their strength is increased to produce the large magnetic
fluxes associated with pores and sunspots, rather than imposing a
given geometry to begin with.

There are very strong horizontal fields, in low lying loops, 
connecting the opposite polarity field concentrations because of the
close proximity of the opposite polarity regions.  This produces
elongated granules and penumbral-like features, but the crowding
does not allow them to grow as in observed sunspots.

\begin{figure}
  \centerline{\includegraphics[width=\textwidth]{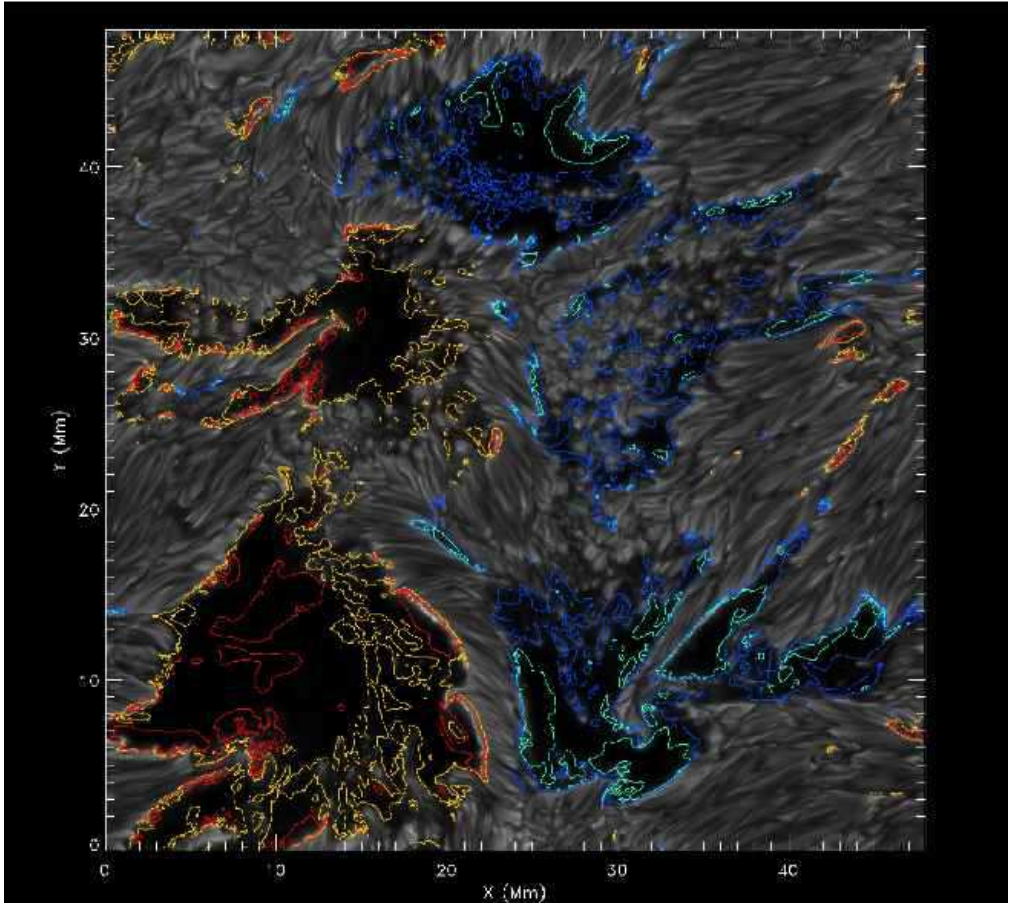}}
  \caption{Emergent continuum intensity image of simulated active region with
  vertical magnetic field contours at $\pm$ 2 (yellow and blue), $\pm$ 2.5 (red and green) kG.
  }
  \label{fig:AR}
\end{figure}

The active region, so far, has evolved for 2 hours.  During this
time the areas of strong vertical field concentrations have maintained
their integrity (Fig.~\ref{fig:AR}: yellow and dark blue contours
are vertical magnetic field $\pm$ 2 kG and red and green are $\pm$
2.5 kG).  Areas with weaker vertical field have gotten filled in
by convective plumes from their periphery plunging into their
interior and dispersing the magnetic field into small, isolated
clumps (See region in the center-right of Fig.~\ref{fig:AR} compared
with the same region in the very center of Fig.~\ref{fig:ARgrowth}
at an earlier time).

\begin{figure}
  \centerline{\includegraphics[width=.7\textwidth]{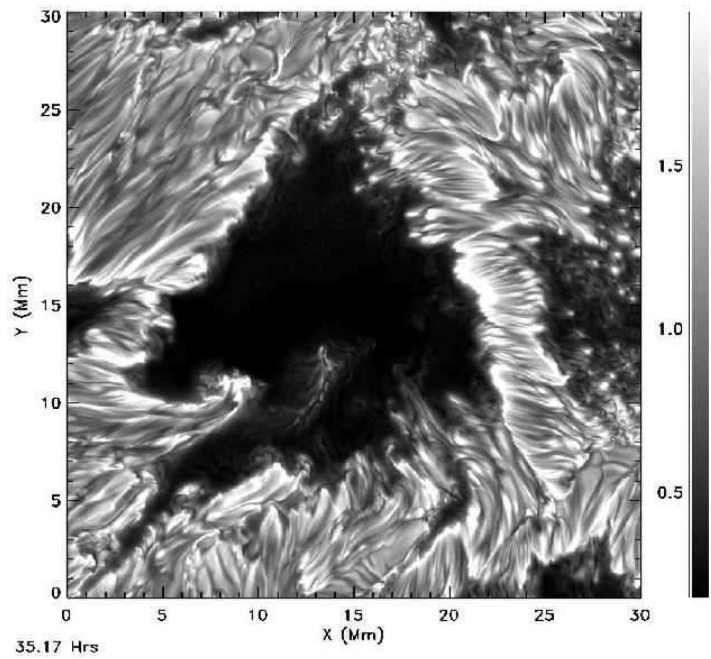}}
  \caption{Spot at lower left corner: emergent continuum intensity I/$<$I$>$.
  }
  \label{fig:spot1}
\end{figure}

\begin{figure}
  \centerline{\includegraphics[width=.7\textwidth]{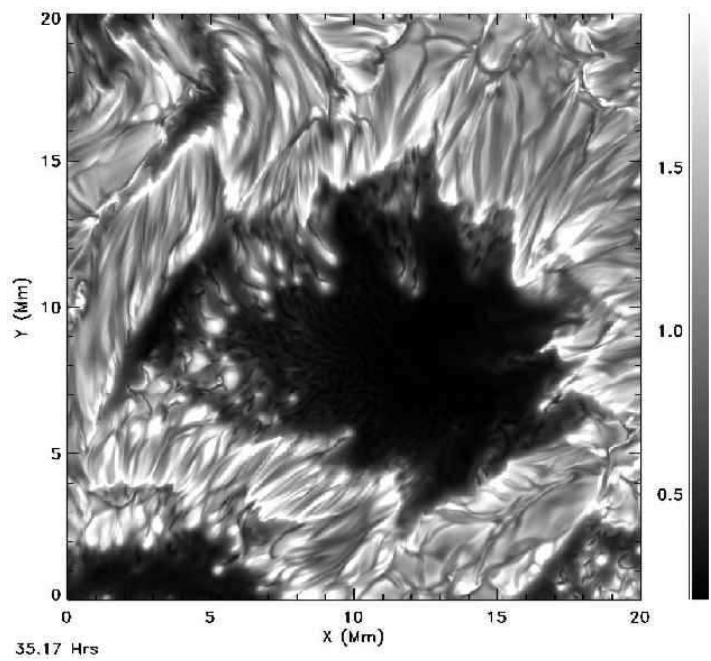}}
  \caption{Spot at top center: emergent continuum intensity I/$<$I$>$.
  }
  \label{fig:spot2}
\end{figure}

Figures \ref{fig:spot1} and \ref{fig:spot2} are more detailed images
of the spots in the lower left and top center of Fig.~\ref{fig:AR},
but at a slightly earlier time.
The short light bridge in Fig.~\ref{fig:spot1} has the dark center
lane and segmented appearance of observed light bridges.  To the
upper left in Fig.~\ref{fig:spot1} there are strong shearing motions
between the elongated granules.  There is a very bright region on
the left edge of the spot in Fig.~\ref{fig:spot1} where there is a
very strong vertical field concentration (see Fig.~\ref{fig:AR}).

\section{Stokes Profiles}

\begin{figure}
  \plotone{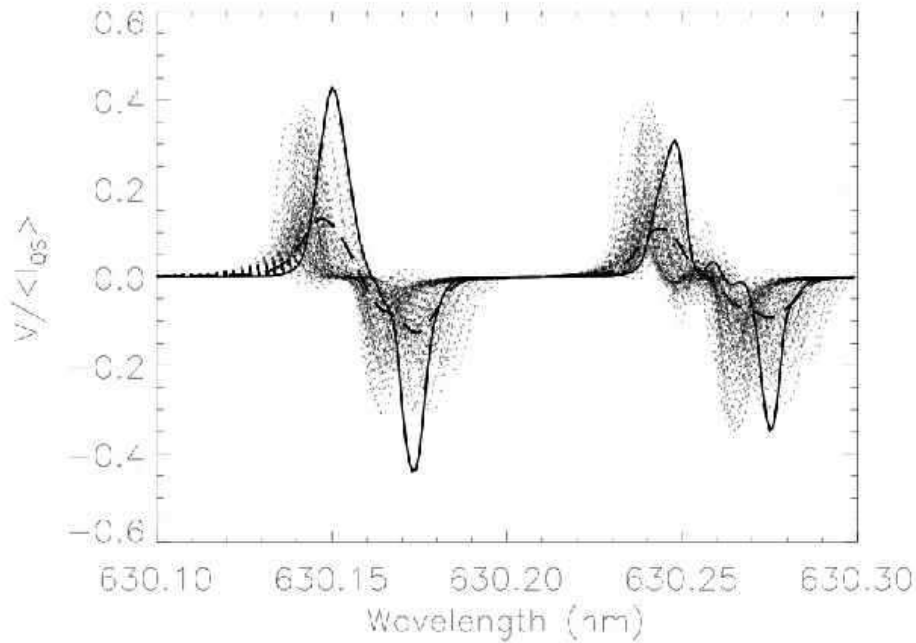}
  \caption{Stokes V profiles for Fe I 630.15 and 630.25 nm lines for
  a penumbral region of a simulated spot:  Original profile (solid
  line), neighboring profiles (dotted lines) and profile after
  Hinode mtf, slit diffraction and frequency smoothing (dashed
  line).
  }
  \label{fig:Vprofile}
\end{figure}

In the centers of the pores/spots, the stokes profiles after passing
through the Hinode optical path (as represented by the 50 cm annular
psf, the slit diffraction and the 24 m{\AA} frequency smoothing)
are identical to the profiles from the individual simulation
pixels (24 km resolution).  However, in the penumbral like features
bordering the pores/spots, where the magnetic field is predominantly
horizontal and the field, thermodynamic and velocity structures
have small scale variations, the modified profiles are significantly
different from individual pixel profiles (Fig.~\ref{fig:Vprofile}).
Raw data, stokes profiles I,Q,U,V and profiles as modified by the 
Hinode optical path are available on line at http://steinr.pa.msu.edu/$\sim$bob/Stokes.

\section{Comparison with Hinode Observations}

\begin{figure}
  \centerline{\includegraphics[width=2.5in]{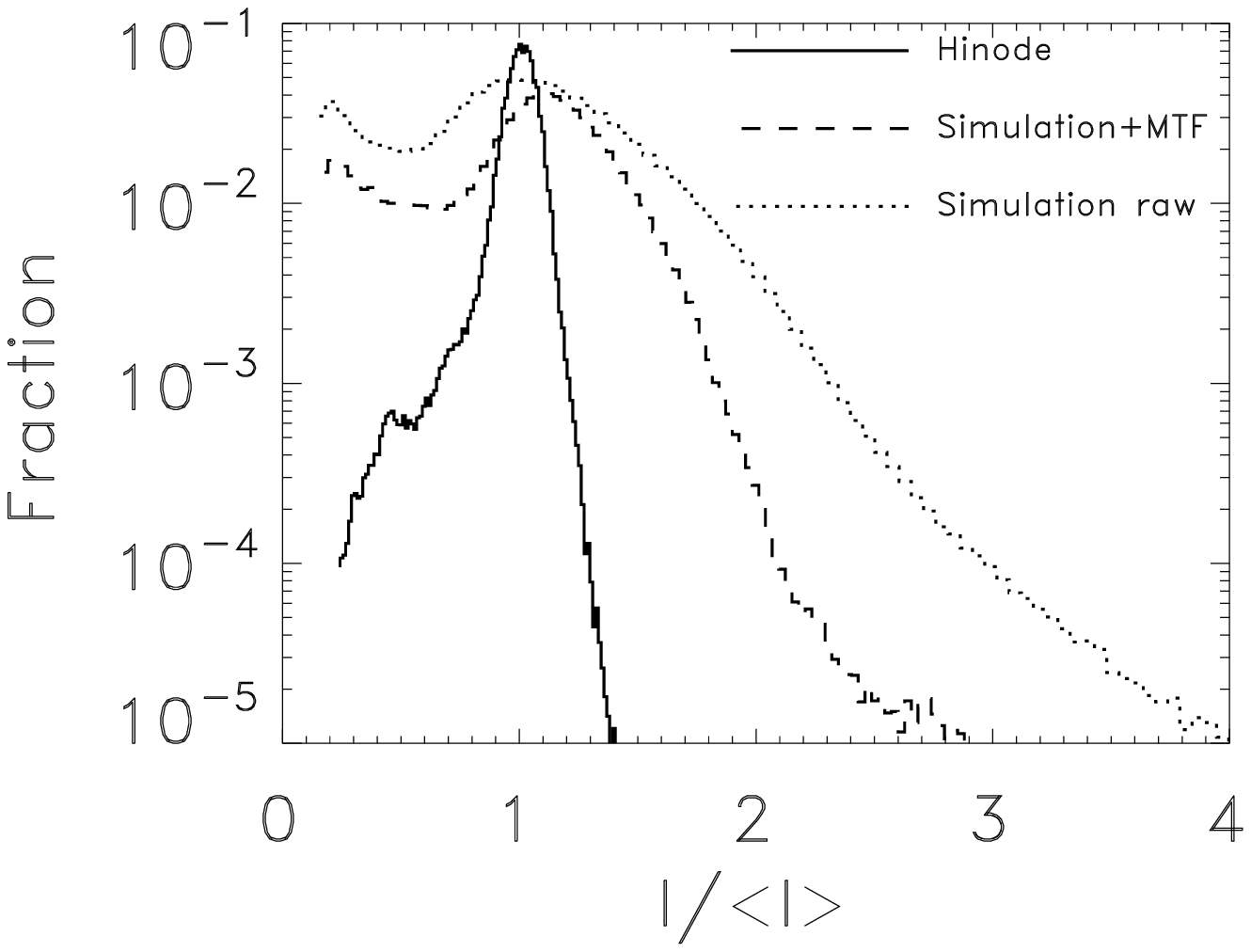}
  \includegraphics[width=2.5in]{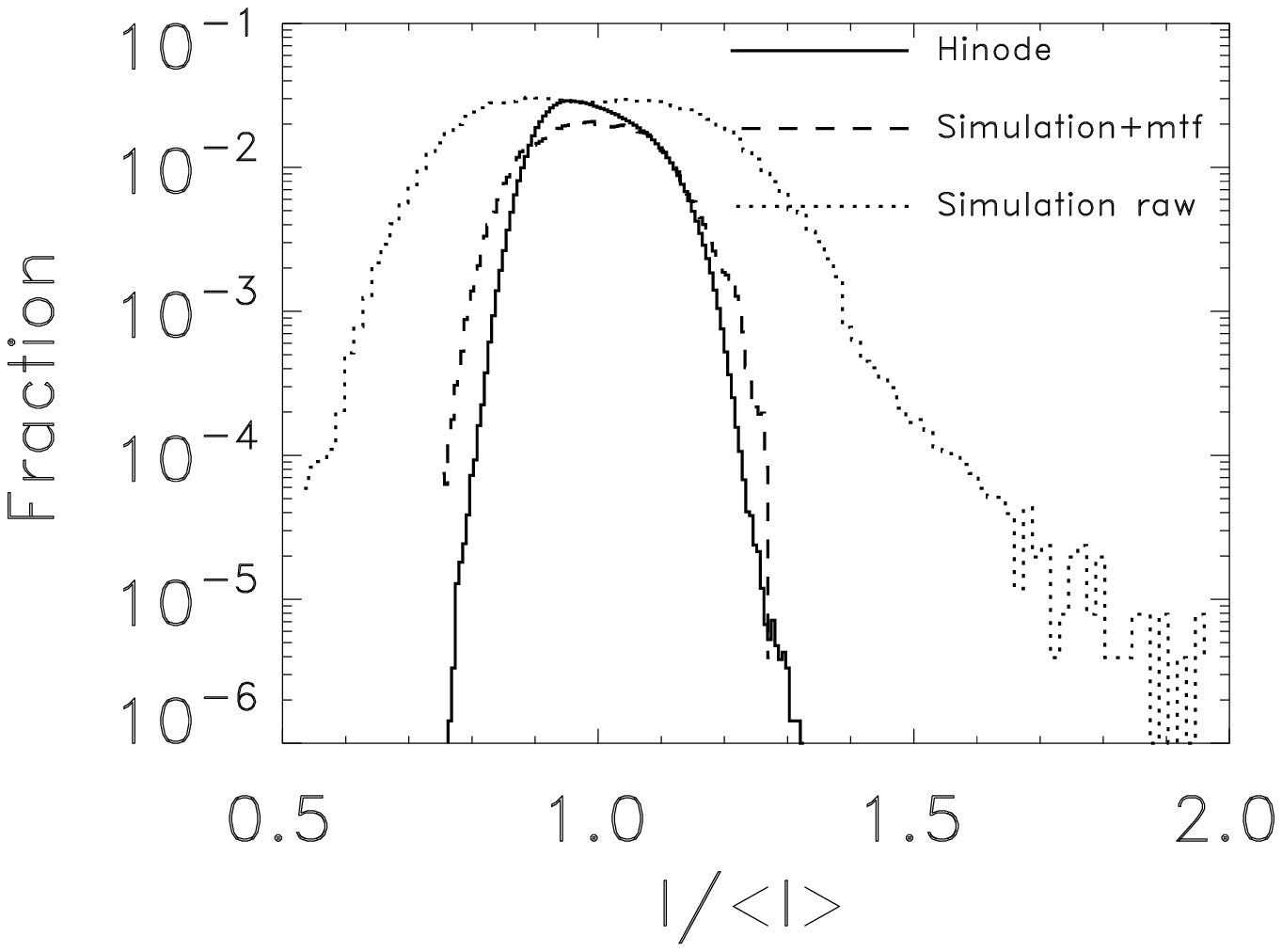}}
  \caption{Emergent continuum intensity distribution, comparing
  simulation results and Hinode observations, for an active 
  region (left) and quiet Sun (right).}
  \label{fig:Int_sim_hinode}
\end{figure}

\begin{figure}
  \centerline{\includegraphics[width=2.5in]{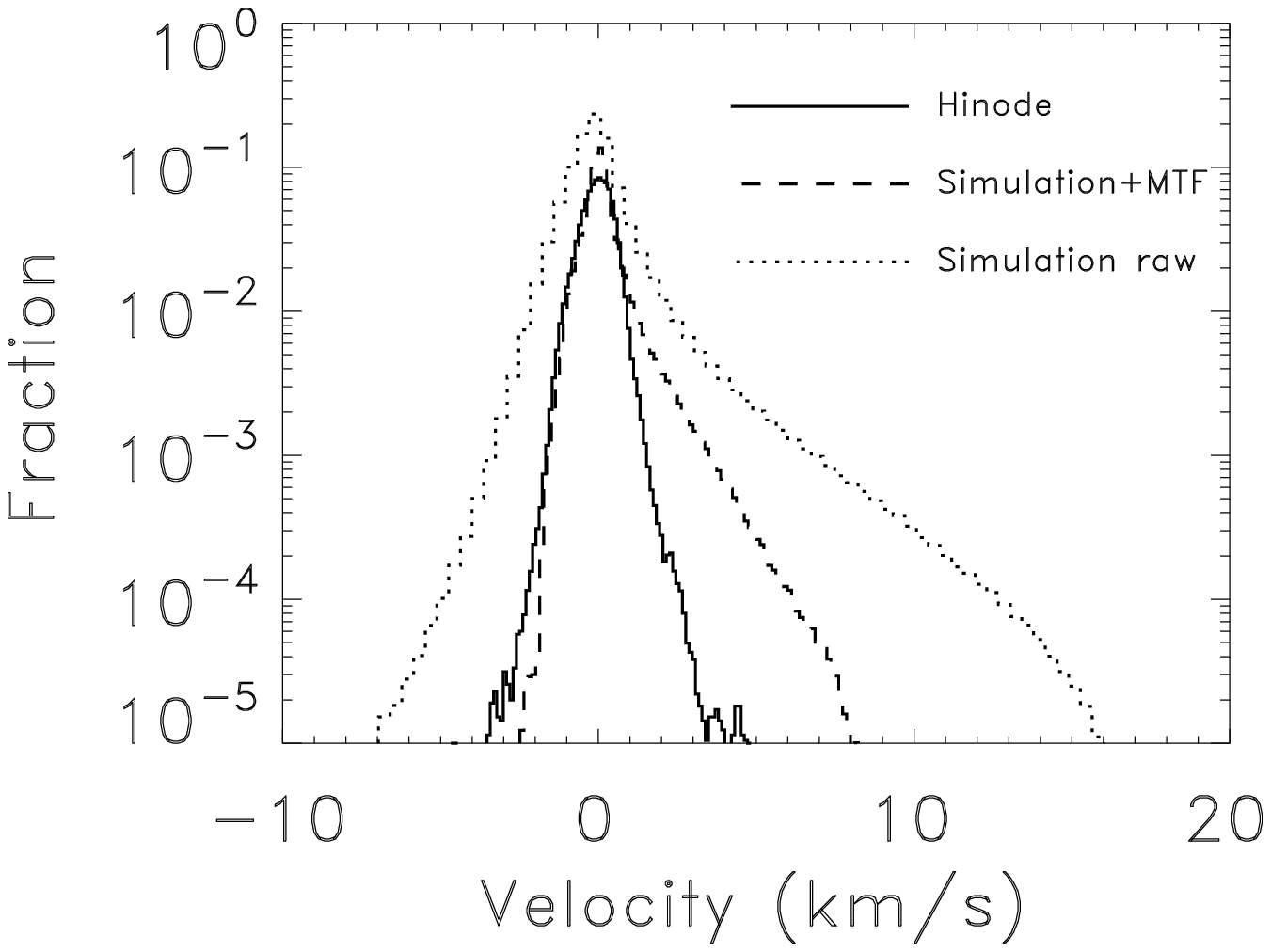}
  \includegraphics[width=2.5in]{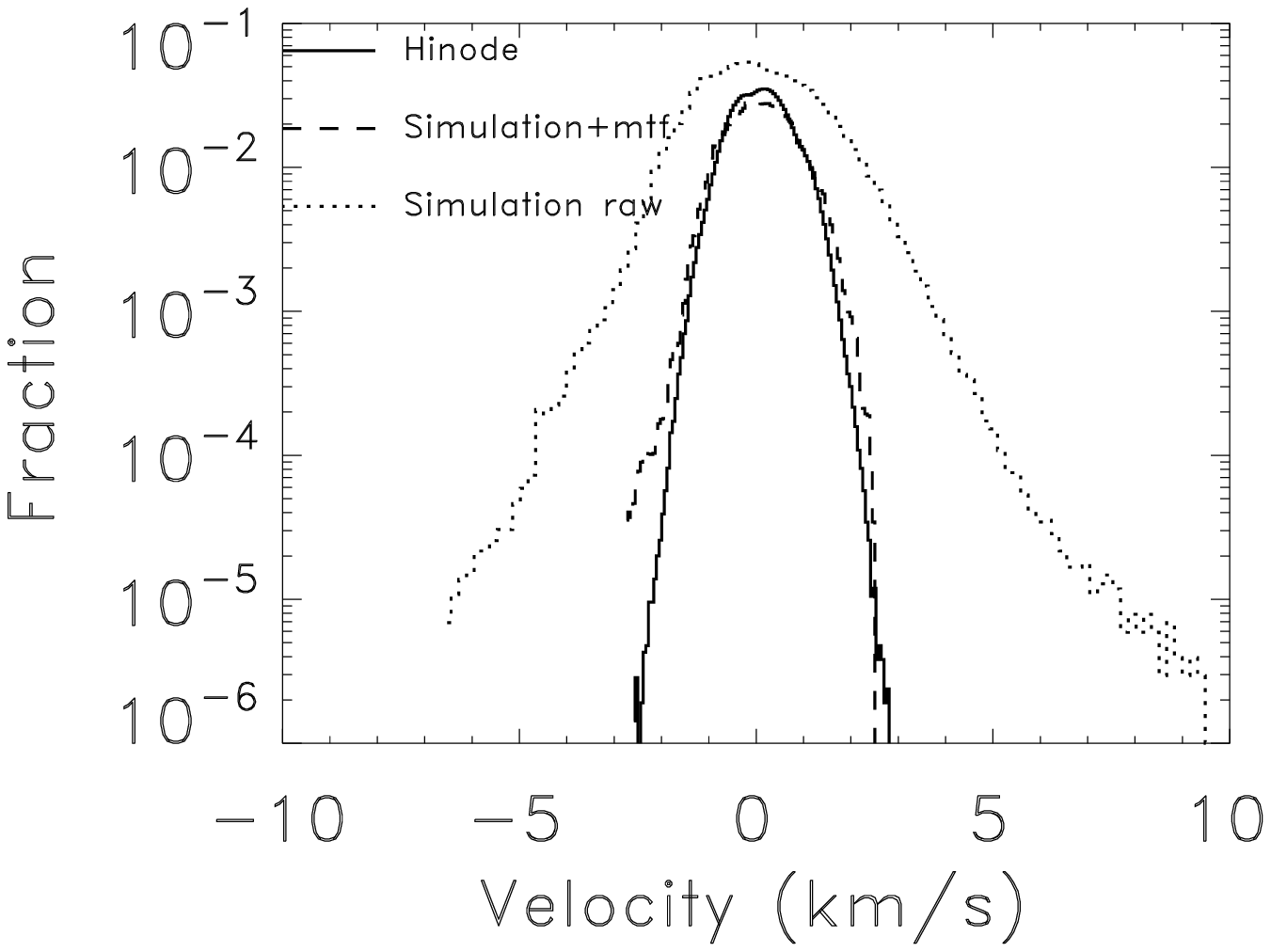}}
  \caption{Vertical velocity distribution, comparing simulation 
  results and Hinode observations, for an active region (left) 
  and quiet Sun (right).  Negative velocities are upward and positive
  velocities are downward.}
  \label{fig:Vel_sim_hinode}
\end{figure}

In addition to the active region simulation described above, we
also have performed a simulation with an initial uniform vertical
magnetic field imposed on a snapshot of hydrodynamic convection.
This calculation is being performed on a domain 12 Mm wide by 6 Mm
deep.  (There is no point to going deeper on such a narrow domain
because the convective cells would become artificially constrained
at larger depths by the periodic horizontal boundary conditions.)
The vertical field was initial 10 G and was increased with a time
scale of 2 hours.  We have compared the emergent continuum intensity
and vertical velocity from both the active region simulation and
this vertical field simulation at the time when the average vertical
field was 60 G, with observations from Hinode that were provided
by Reza Rezaei (Figs.~\ref{fig:Int_sim_hinode} and \ref{fig:Vel_sim_hinode}).

For the weak vertical field unipolar region compared to the quiet
Sun observations, the raw simulation distribution of both intensity
and velocity is much wider than observed.  However, when the Hinode
MTF is applied to the simulation results there is excellent agreement
with the observations.  For the active region simulation, on the
other hand, agreement is not so good.  The simulation results have
much larger downward velocities and a larger fraction of both high
and low intensities.  The larger fraction of low intensities is due
to the crowding of the dark spot-like areas.  The origin of the
much brighter locations is unclear.  Figs.~\ref{fig:spot1} and
\ref{fig:spot2} show some very bright edges to the magnetic
concentrations where the horizontal field turns over into vertical
field.

\acknowledgements This work was funded by NSF grant AST 0605738 and
NASA grants NNX07AO71G, NNX07AH79G and NNX08AH44G.  The calculations
were performed on the Pleiades supercomputer of the NASA Advanced
Supercomputing Division.  It would not have been possible without
this support.

\bibliography{steinh4}

\end{document}